\documentclass[twocolumn]{aastex6}
\usepackage{graphicx}
\usepackage[flushleft]{threeparttable}
\shorttitle{Kiloparsec--scale image of a merger at $z\approx6.2$}
\shortauthors{Decarli et al.}

\def\Lsun{L$_\odot$}
\def\Msun{M$_\odot$}

\def\Cii{[C\,{\sc ii}]}

\def\kms{km\,s$^{-1}$}

\def\lsim{\mathrel{\rlap{\lower 3pt \hbox{$\sim$}} \raise 2.0pt \hbox{$<$}}}
\def\gsim{\mathrel{\rlap{\lower 3pt \hbox{$\sim$}} \raise 2.0pt \hbox{$>$}}}

\begin{document}

\title{
ALMA and HST kiloparsec--scale imaging of a quasar--galaxy merger at $z\approx 6.2$
}

\author{
Roberto Decarli\altaffilmark{1},
Massimo Dotti\altaffilmark{2,3},
Eduardo Ba\~{n}ados\altaffilmark{4,5},
Emanuele Paolo Farina\altaffilmark{4,6},
Fabian Walter\altaffilmark{4,7,8},
Chris Carilli\altaffilmark{7,9}, 
Xiaohui Fan\altaffilmark{10}, 
Chiara Mazzucchelli\altaffilmark{11}, 
Marcel Neeleman\altaffilmark{4},
Mladen Novak\altaffilmark{4},
Dominik Riechers\altaffilmark{12}, 
Michael A. Strauss\altaffilmark{13}, 
Bram P.~Venemans\altaffilmark{4},
Yujin Yang\altaffilmark{14},
Ran Wang\altaffilmark{15} 
}
\altaffiltext{1}{INAF -- Osservatorio di Astrofisica e Scienza dello Spazio di Bologna, via Gobetti 93/3, I-40129, Bologna, Italy. E-mail: {\sf roberto.decarli@inaf.it}}
\altaffiltext{2}{Universit\`{a} degli Studi di Milano--Bicocca, Dipartimento di Fisica G.~Occhialini, piazza della Scienza 3, I-20126 Milano, Italy}
\altaffiltext{3}{INFN, Sezione Milano--Bicocca, Piazza della Scienza 3, I-20126 Milano, Italy}
\altaffiltext{4}{Max-Planck Institut f\"{u}r Astronomie, K\"{o}nigstuhl 17, D-69117, Heidelberg, Germany}
\altaffiltext{5}{The Observatories of the Carnegie Institution for Science, 813 Santa Barbara St., Pasadena, CA 91101, USA}
\altaffiltext{6}{Max-Planck-Institut f\"{u}r Astrophysik, Karl-Schwarzschild-Str. 1, D-85748 Garching, Germany}
\altaffiltext{7}{National Radio Astronomy Observatory, Pete V.\,Domenici Array Science Center, P.O.\, Box O, Socorro, NM, 87801, USA}
\altaffiltext{8}{Astronomy Department, California Institute of Technology, MC249-17, Pasadena, California 91125, USA}
\altaffiltext{9}{Battcock Centre for Experimental Astrophysics, Cavendish Laboratory, University of Cambridge, 19 J J Thomson Avenue, Cambridge CB3 0HE, UK}
\altaffiltext{10}{Steward Observatory, University of Arizona, 933 N. Cherry St., Tucson, AZ  85721, USA}
\altaffiltext{11}{European Southern Observatory, Alonso de C\'{o}rdova 3107, Vitacura, Regi\'{o}n Metropolitana, Chile}
\altaffiltext{12}{Cornell University, 220 Space Sciences Building, Ithaca, NY 14853, USA}
\altaffiltext{13}{Department of Astrophysical Sciences, Princeton University, Princeton, New Jersey 08544, USA}
\altaffiltext{14}{Korea Astronomy and Space Science Institute, Daedeokdae-ro 776, Yuseong-gu Daejeon 34055, South Korea}
\altaffiltext{15}{Kavli Institute of Astronomy and Astrophysics at Peking University, 5 Yiheyuan Road, Haidian District, Beijing 100871, China}

\begin{abstract}
We present kpc--scale ALMA and {\em HST} imaging of the quasar PJ308--21 at $z$=$6.2342$, tracing dust, gas (via the \Cii{} 158\,$\mu$m line) and young stars. At a resolution of $\sim0.3''$ ($\approx1.7$\,kpc), the system is resolved over $>4''$ ($>$20\,kpc). In particular, it features a main component, identified to be the quasar host galaxy, centered on the accreting supermassive black hole; and two other extended components on the West and East side, one redshifted and the other blueshifted relative to the quasar. The \Cii{} emission of the entire system stretches over $>$1500\,\kms{} along the line of sight. All the components of the system are observed in dust, \Cii{}, and rest--frame UV emission. The inferred \Cii{} luminosities [(0.9--4.6)$\times 10^9$\,\Lsun], dust luminosities [(0.15--2.6)$\times10^{12}$\,\Lsun], and rest--frame UV luminosities [(6.6--15)$\times10^{10}$\,\Lsun], their ratios, and the implied gas/dust masses and star formation rates [11--290\,\Msun\,yr$^{-1}$] are typical of high--redshift star--forming galaxies. A toy model of a single satellite galaxy that is tidally stripped by the interaction with the quasar host galaxy can account for the observed velocity and spatial extent of the two extended components. An outflow interpretation of the unique features in PJ308--21 is not supported by the data. PJ308--21 is thus one of the earliest galaxy mergers imaged at cosmic dawn. 
\end{abstract} 

\section{Introduction} 

Quasars are the most luminous non-transient sources in the universe. Their enormous energy output, powered by intense ({\bf $\gsim$5\,\Msun{}\,yr$^{-1}$}) and radiatively efficient gas accretion onto a supermassive ($10^{8-10}$\,\Msun) black hole makes them ideal laboratories to study the intergalactic medium and the ionization history of the early universe \citep[e.g.,][]{banados18,davies18}, the build-up of massive black holes \citep[e.g.,][]{volonteri12,mazzucchelli17}, the formation of the first massive galaxies \citep[e.g.,][]{venemans17}, and the development of the first large--scale structures in the universe \citep[e.g.,][]{balmaverde17}.

Models of early massive black hole formation postulate that $z>6$ quasars reside in the extreme peaks of the large--scale density structure \citep[e.g.,][]{angulo12}, where gravitational interactions and mergers are expected. Direct observational evidence of these processes is challenging at these redshifts. With only few exceptions \citep[see, e.g.,][]{farina17}, companion sources are faint and often identified only via broad--band imaging \citep[e.g.,][]{stiavelli05,mcgreer14}, thus leaving room for contamination by foreground sources. 

The exceptional sensitivity and imaging power of the Atacama Large Millimeter/sub-millimeter Array, ALMA, now allows us to image the dust and cold gas reservoirs (the latter probed in particular via the \Cii{} 158\,$\mu$m line) of galaxies in the early universe in detail. \citet{decarli17} used ALMA to identify four \Cii{}--bright galaxies in close proximity to $z>6$ quasars, out of a survey of 27 objects \citep{decarli18}. Two of these systems, PJ308--21 and PJ231--20, show projected separations between the quasar and the companion galaxy of $\lsim$10\,kpc, thus making an on--going merger scenario very plausible. The quasar PJ167--13 also shows a very close \Cii{}--emitting companion (\citealt{willott17}, Neeleman et al.~in prep.).
Similar cases have been found also at lower redshifts \citep[e.g.,][]{trakhtenbrot17,diazsantos18} as well as in the proximity of Lyman-Break Galaxies at $z\approx6$ \citep{jones17}.
%In particular, the sensitive high--resolution ALMA imaging of the quasar W2246--0526 shows spectactular tidal features which are the smoking gun of an on--going galaxy merger \citep{diazsantos18}. % The only other clearcut example of close galaxy interaction revealed by ALMA at $z\gsim 6$ is the system associated with the lensed galaxy XXX studied by \citet{marrone18}.

In this work, we present new high--resolution ALMA and {\em Hubble Space Telescope} ({\em HST}) imaging of the quasar+companion system  PSO J308.0416--21.2339 (\citealt{banados16}; hereafter PJ308--21) at $z$=$6.2342$. The synergy of ALMA and {\em HST} observations reveals the morphology and internal dynamics of this system, and properties of its star--forming medium.

Throughout the paper we assume a standard $\Lambda$CDM cosmology with $H_0=70$ km s$^{-1}$ Mpc$^{-1}$, $\Omega_{\rm m}=0.3$ and $\Omega_{\Lambda}=0.7$ \citep[consistent with the measurements by the][]{planck15}. In this framework, at $z=6.2342$ the luminosity distance is 60,366 Mpc, and $1''$ on sky corresponds to a projected physical separation of $5.59$\,kpc. Magnitudes are reported in the AB photometric system.

\section{Observations and data reduction}\label{sec_observations}

\subsection{ALMA}

The ALMA observations of PJ308--21 discussed here include the original low--resolution ($\sim1''$) data from the survey of \Cii{} and underlying dust continuum in $z>5.94$ quasars by \citet{decarli18} and \citet{venemans18} (program ID: 2015.1.01115.S). In addition, we present new follow--up observations at high resolution ($\sim 0.3''$) obtained in a Director's Discretional Time allocation (program ID: 2016.A.00018.S). These high--resolution observations were collected in two executions on May 3 and 5, 2017, while the array was in C40-5 configuration. We adopted the same frequency setting and pointing direction as for the low--resolution data, thus encompassing the redshifted \Cii{} line ($\nu_0=1900.547$\,GHz) at the frequency of 263.18\,GHz. The high--resolution observations include $\sim1$\,hr of on-source data. The quasars J1924-2914, J2056-4714, J2042-2255 served as bandpass, flux, and phase calibrator, respectively.

\begin{figure*}
  \includegraphics[width=0.49\textwidth]{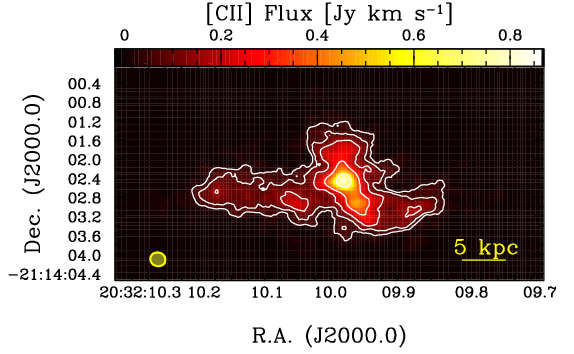}
  \includegraphics[width=0.49\textwidth]{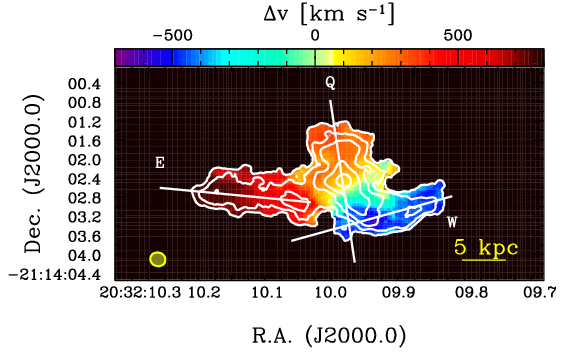}\\
  \includegraphics[width=0.49\textwidth]{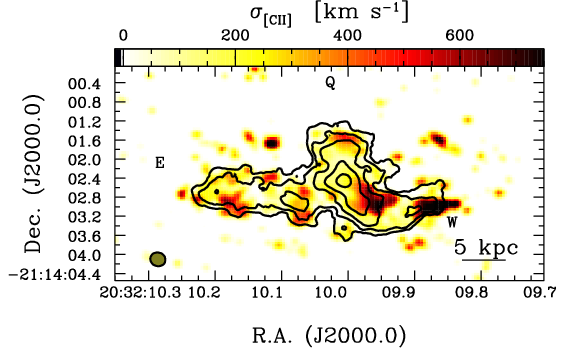}
  \includegraphics[width=0.49\textwidth]{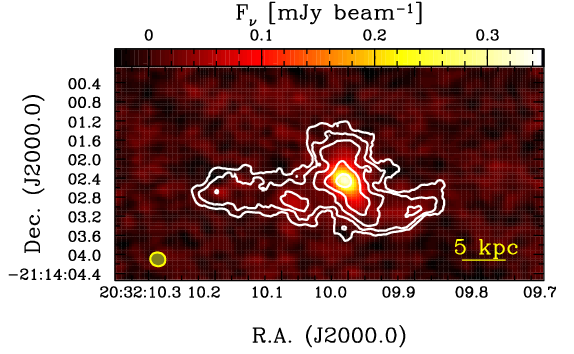}\\
\vspace{-5mm}
\caption{ALMA imaging of the \Cii{} 158\,$\mu$m and underlying dust continuum emission in PJ308--21. Moment zero ({\em top left panel}), one ({\em top right panel}), and two ({\em bottom left panel}) maps of the (continuum--subtracted) \Cii{} line. The bulk of the \Cii{} emission arises along the North--South direction, spatially consistent with the quasar (labeled Q). Additionally, two blobs on the western and eastern sides are also apparent (labeled W and E). A similar morphology is apparent in the dust continuum emission (bottom panel). The \Cii{} intensity contours (at 0.04, 0.08, 0.16, 0.32, 0.64\,Jy\,\kms{}\,beam$^{-1}$), the beam size, and the equivalent physical scale are shown for comparison in all panels. }
\label{fig_alma_maps}
\end{figure*}

\begin{figure}
  \includegraphics[width=0.49\textwidth]{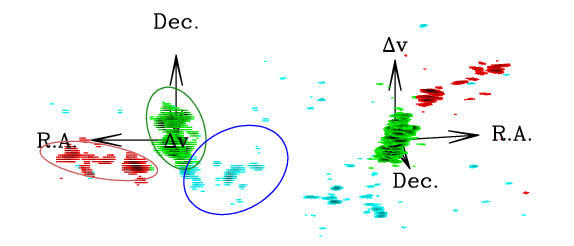}\\
  \includegraphics[width=0.49\textwidth]{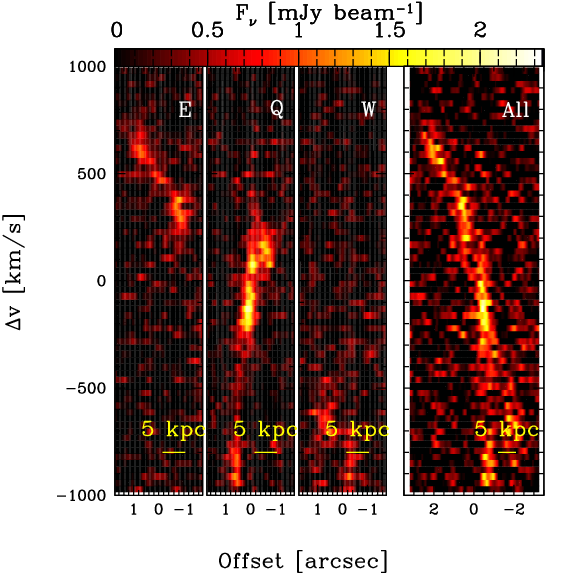}\\
\vspace{-5mm}
\caption{Gas velocity structure in PJ308--21, as mapped by ALMA imaging of the \Cii{} 158\,$\mu$m line. {\em Top:} 3D (R.A., Dec., $\Delta v$) renderings of the continuum--subtracted \Cii{} emission (the first one being equivalent to the \Cii{} map shown on the left), showing the complex morphology and dynamics in the system.  The three components of PJ308--21 are marked in green, cyan, and red. {\em Bottom:} Position--velocity diagrams extracted along the main axes of each component (shown as bars in the velocity field map of Fig.~\ref{fig_alma_maps}), as well as from a $3''$ wide slit along the West--East direction that encapsulates the entire system.}
\label{fig_alma_vel}
\end{figure}

%The whole structure structure extends over $>$20\,kpc along the East--West direction.  The velocity field reveals that the Western and Eastern clouds are blue-- and red--shifted compared to the quasar host, by $\gsim$500\,\kms{}. The quasar host shows a clear velocity gradient along the North--South direction. 

We ran the ALMA pipeline in Common Astronomy Software Applications, CASA \citep[version 4.7.2;][]{mcmullin07} for data calibration. Typical phase rms is $<$20$^\circ$ even at the longest baselines ($\approx 1.1$\,km). We concatenated the low-- and high--resolution datasets, and inverted the visibilities using the task \textsf{tclean}. We created a datacube by adopting Briggs weighting with robustness parameter set to 2 (i.e., `natural' visibility weighting). The synthesized beam size is $0.38''\times 0.30''$ (at Position Angle=$82^\circ$). We sampled the spectral dimension in 30\,\kms{} wide channels. The typical rms of the noise is 0.20\,mJy\,beam$^{-1}$ per 30\,\kms{} channel. Following \citet{decarli18}, we also create a line--free continuum image, which is then used to perform continuum--subtraction via the task \textsf{uvsub}. Because of the intricated velocity structure of this system, spanning a large range of frequency, we capitalize on the line--free channels of the full available spectral coverage in the creation of the continuum image, which reaches a rms of $10.6$ $\mu$Jy\,beam$^{-1}$. The continuum--subtracted \Cii{} cube is then collapsed along the frequency axis, after applying a S/N$>$1.5 mask, in order to create moment 0 and 1 maps. Fig.~\ref{fig_alma_maps}, shows these two maps, together with the dust continuum map. We extract \Cii{} position--velocity diagrams along various directions. Finally, we produce 3D renderings of the continuum--subtracted \Cii{} emission, to fully capture the complex morphological and kinematical structure of the system (see Fig.~\ref{fig_alma_vel}).

\subsection{\it HST}

The {\it HST} observed PJ308--21 on May 4, 2017 (program 14876), using the F140W filter on the Wide Field Camera 3 (WFC3) IR arm. At $z=6.2342$, the pivot wavelength of the filter ($\lambda=1.392$\,\AA{}) samples the rest-frame far UV (1925\,\AA{}) emission. The total integration was 2611.75\,s, split into 4 frames with small dithering offsets (as in the WFC3-IR-DITHER-BOX-MIN template). 

The data reduction was performed using the standard {\it HST} pipeline, in particular the \textsf{AstroDrizzle} package (version 2.1.3.dev). The pixel scale of the final image is 0.128\,$''$\,pixel$^{-1}$. We reach a 5-$\sigma$ surface brightness limit of 26.6\,mag\,arcsec$^{-2}$ for a 1 arcsec$^2$ aperture. 
% or 1.34 nJy\,arcsec$^{-2}$

\begin{figure}
\includegraphics[width=0.23\textwidth]{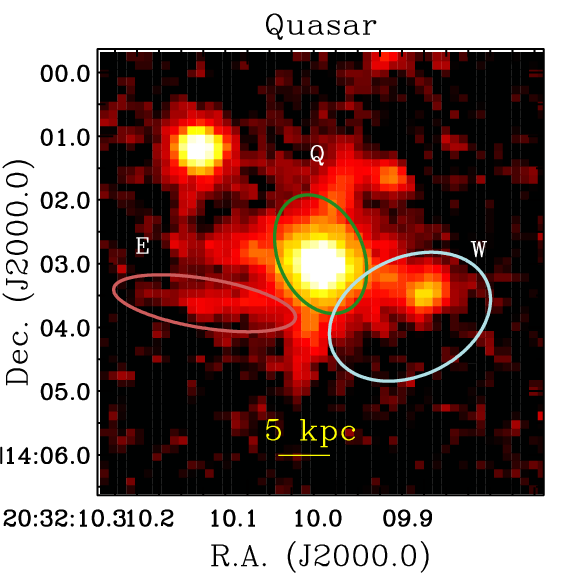}
\includegraphics[width=0.23\textwidth]{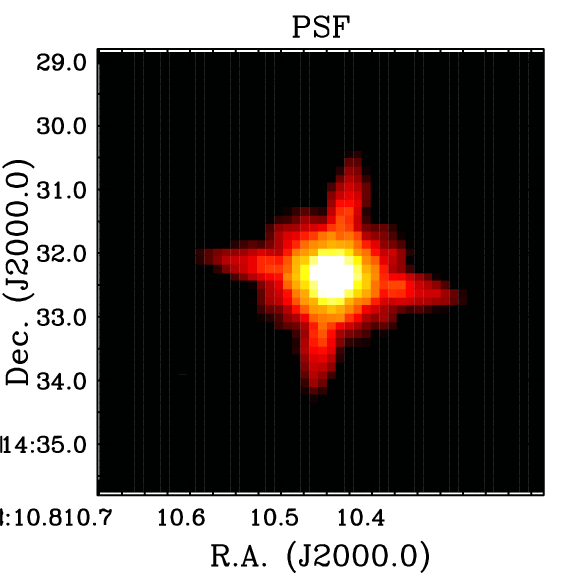}\\
\includegraphics[width=0.48\textwidth]{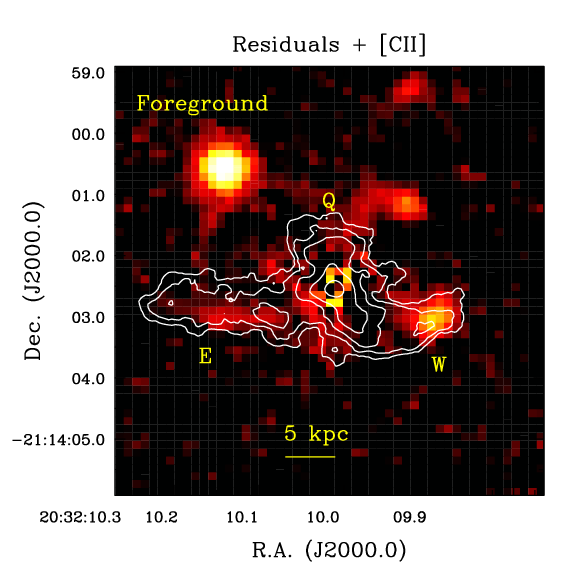}\\
\caption{{\em HST} imaging of PJ308--21. {\em Top:} The observed quasar + host galaxy emission (left) and the PSF model (right). The ellipses mark the apertures used to define the three components discussed in this analysis. {\em Bottom:} Residuals after PSF subtraction. We observe extended emission stretching well beyond the tails of the PSF. A comparison with the \Cii{} map (contours from Fig.~\ref{fig_alma_maps}) reveals spatial coincidence between the \Cii{} and the starlight emission identified with {\em HST} eastwards and westward of the quasar. An additional blob, located north--west to the quasar in the {\em HST} image, is also likely associated with the system. On the contrary, the bright source North--East of the quasar is a foreground source (see Farina et al.~in prep).}
\label{fig_hst_maps}
\end{figure}

In order to search for extended emission from the host galaxy of PJ308--21 and from the companion source in the {\em HST} image, we model and remove the point-like emission from the quasar. We do so using \textsf{GALFIT} \citep{peng06}, combined with a suite of custom IRAF\footnote{IRAF is distributed by the National Optical Astronomy Observatory, which is operated by the Association of Universities for Research in Astronomy (AURA) under a cooperative agreement with the National Science Foundation. }--based tasks (see, e.g., \citealt{decarli12}). We create a model of the Point Spread Function (PSF) by median--averaging the normalized images of 8 stars in the field, chosen for being distant from potential contaminants and with a flux 1.5--15$\times$ brighter than the quasar (in order to measure the PSF wings well). 
We do not down--select reference stars by their spectral type, thus the PSF model might carry systematic uncertainties due to the color dependence of the empirical PSFs used in the analysis. However, our empirical PSF model appears to work well in subtracting the color-sensitive diffraction spikes. 
We fit this empirical PSF model to the quasar image by allowing the PSF centroid to move by $<1$ pixel, i.e., the PSF is scaled to capture only the nuclear emission. The observed image, the model PSF, and the residuals after PSF subtraction are shown in Fig.~\ref{fig_hst_maps}.

\begin{table}
\caption{\rm Summary of the observed properties of the quasar host (Q) and the eastern (E) and western (W) components. Errors in the measured fluxes are of the order of 10\%, and are most sensitive on the exact shape of the adopted apertures.} \label{tab_obs}
\begin{center}
\begin{tabular}{c|ccc}
\hline
                                                           & Q       & E       & W	 \\
\hline
$\Delta v$              [\kms]  			   & --      & +500    & -750    \\
$F_{\rm [CII]}$         [Jy\,\kms{}]			   & $4.7$   & $1.7$   & $0.9$	 \\
$F_{\nu}$(158$\mu$m)    [mJy]				   & $1.18$  & $0.25$  & $0.13$	 \\
F140W                   [mag]				   & $25.33$ & $25.17$ & $24.44$ \\
\hline
log $L_{\rm [CII]}$     [\Lsun] 			   &  $9.67$ & $ 9.23$ & $ 8.96$ \\
log $L_{\rm IR}$($T$=35\,K) [\Lsun] 			   & $12.12$ & $11.44$ & $11.17$ \\
log $L_{\rm UV}$        [\Lsun] 			   &$>10.82$ & $10.88$ & $11.17$ \\
log $L_{\rm [CII]}/L_{\rm IR}$                             & $-2.46$ & $-2.21$ & $-2.21$ \\
log IRX                 				   &  $1.31$ & $ 0.56$ & $ 0.00$ \\
\hline
log $M_{\rm gas}^{\rm min}$ [\Msun]                        & $10.67$ & $10.24$ & $ 9.97$ \\
log $M_{\rm gas}$ [\Msun]                                  & $11.14$ & $10.71$ & $10.43$ \\
log $M_{\rm dust}$ [\Msun]                                 & $ 8.15$ & $ 7.32$ & $ 7.18$ \\
log SFR$_{\rm UV}$        [\Msun{}\,yr$^{-1}$]		   & $>1.05$ & $ 1.12$ & $ 1.41$ \\
log SFR$_{\rm IR}$        [\Msun{}\,yr$^{-1}$]		   &  $2.30$ & $ 1.62$ & $ 1.34$ \\
log $\Sigma_{\rm SFR,UV}$ [\Msun{}\,yr$^{-1}$\,kpc$^{-2}$] &$>-1.11$ & $-0.90$ & $-0.47$ \\
log $\Sigma_{\rm SFR,IR}$ [\Msun{}\,yr$^{-1}$\,kpc$^{-2}$] & $ 0.14$ & $-0.40$ & $-0.53$ \\
\hline
\end{tabular}
\end{center}
\end{table}

\section{Analysis and results}\label{sec_results}

\subsection{Morphology of the system}

Both the ALMA and {\em HST} images of the PJ308--21 system reveal extended structures. The bulk of the gas, traced by the \Cii{} emission, is organized in a 4.5\,kpc long structure roughly aligned with the North--South direction, and spatially coincident with the quasar emission seen at optical/NIR wavelengths. We will refer to this component as the quasar host galaxy, Q. Its integrated \Cii{} flux is 4.7\,Jy\,\kms{}, corresponding to a \Cii{} luminosity of $L_{\rm [CII]}=4.6\times10^9$\,\Lsun{}\footnote{This is $\sim 25$\% higher than the one originally published in \citet{decarli17} because the superior depth of the new data allows us to better capture the full extent of the emission.}. The {\em HST} image of the host is mostly outshone by the quasar emission, but residuals of modest instensity (to a flux density of $\sim 25.3$\,mag) are apparent in the South--East after PSF subtraction.

The \Cii{} emission originally identified as a quasar companion \citep{decarli17} extends $>$10\,kpc eastward of the quasar host (hereafter, E). It shows a marked velocity gradient along the east--west axis, with a shift with respect to the quasar host rest frame that rises from $+300$\,\kms{} to about $+700$\,\kms{} at increasing projected distance. The \Cii{} flux of the Eastern cloud is 1.7\,Jy\,\kms{}, yielding a \Cii{} luminosity of $L_{\rm [CII]}=1.7\times10^9$\,\Lsun{}. The \Cii{} emission also extends westward of the quasar (W). In this case the velocity difference drops at larger projected distance, ranging from $-900$\,\kms{} at the southern edge of the quasar host galaxy to $-600$\,\kms{} further away. The integrated \Cii{} flux of this component is 0.9\,Jy\,\kms{}, corresponding to a luminosity of $L_{\rm [CII]}=9.0\times10^8$\,\Lsun{}. The dust emission shows a similar morphology (see Fig.~\ref{fig_alma_maps}).

Remarkably, the \Cii{} emission in both E and W is spatially aligned with and shows a similar morphology to the diffuse emission detected with {\em HST}. This unambiguously associates the latter with young stars at the redshift of the quasar, it is not due to a projected foreground object. An additional blob is observed north--west of the quasar, and appears to be connected to the quasar host; however, because of the lack of a clear \Cii{} counterpart, we cannot rule out that it is a foreground source. Finally, a relatively bright object located $\sim2.5$'' north--east of PJ308--21 is identified with a foreground source (see Farina et al.~in prep), visible also in the Pan--STARRS $i$ and $z$ bands (with measured fluxes of $22.81\pm0.11$ and $22.15\pm0.17$, respectively; \citealt{chambers16}).

% F140W = 25.231 mag in the NW source

\subsection{Gas and dust masses, ISM properties}

The most common tracer of (molecular) gas in high--redshift galaxies is carbon monoxide, CO \citep[for a review, see][]{carilli13}, which has not been observed in PJ308--21 yet. Instead, we infer order-of-magnitude constraints on the gas mass budget from first principles on the \Cii{} emissivity, following \citet{venemans17}. We use the observed \Cii{} line emission to infer the mass in singly--ionized carbon, $M_{\rm C+}$, under the assumptions that the \Cii{} emission is optically--thin and that ionized carbon is in local thermodynamical equilibrium:
\begin{equation}\label{eq_mass_cii}
\frac{M_{\rm C+}}{\rm M_\odot}=2.92\times 10^{-4}\,\frac{Q(T_{\rm ex})}{4} e^{91.2/T_{\rm ex}}\,\frac{L'_{\rm [CII]}}{\rm K\,km\,s^{-1}\,pc^2}
\end{equation}
where $Q(T_{\rm ex})=2+4 \exp(-91.2/T_{\rm ex})$ is the partition function, and $T_{\rm ex}$ is the excitation temperature. For a typical photon--dominated region value of $T_{\rm ex}$=100\,K \citep[see, e.g.,][]{meijerink07,venemans17}, we infer $M_{\rm C+}$ = $(13.9,\,5.1,\,2.7)\times10^6$\,\Msun{} for Q, E, and W, respectively. We then derive an associated gas mass $M_{\rm gas}^{\rm min}$, by assuming the proto--solar carbon abundance (C/H=$2.95\times 10^{-4}$, \citealt{asplund09}). This yields $M_{\rm gas}^{\rm min}$=$(4.7,\,1.7,\,0.9)\times 10^{10}$\,\Msun{} for Q, E, and W, respectively. These are lower limits in that the estimated gas mass would increase if we correct for the carbon that is not in singly--ionized form, if we allow for a lower metallicity, or if we account for suppressed \Cii{} emission due to collisional de-excitation, non-negligible optical depth, etc. We stress however that the system (or parts of it) might not be in thermodynamical equilibrium, thus invalidating our $M_{\rm gas}^{\rm lim}$ estimates. E.g., shocks can enhance \Cii{} emission \citep[see, e.g.,][]{appleton13}. Alternatively, \citet{zanella18} proposed the use of \Cii{} as a tracer of the molecular gas mass, via an empirically--calibrated \Cii{}--to--H$_2$ mass ratio $\alpha_{\rm [CII]}=30$\,\Msun{}/\Lsun{}. In adopting a fixed $\alpha_{\rm [CII]}$, one should keep in mind that a plethora of physical processes (intensity and hardness of the radiation field, collisional de-excitation, extent and intensity of the starburst event, optical depth, etc) might alter the emerging intensity of the \Cii{} for a given gas mass. This yields \Cii{}--based gas masses of $M_{\rm gas}=(13.9, 5.1, 2.7)\times 10^{10}$\,\Msun{}, for the Q, E, and W, respectively -- roughly three times larger than our $M_{\rm gas}^{\rm lim}$ lower limits derived from first principles. 

From the dust continuum images, we measure continuum flux densities of $F_{\rm 1.1mm}$=1.01, 0.19, and 0.13 mJy for Q, E, and W respectively. Assuming that the dust emission can be described by a modified black body with fixed dust emissivity $\beta=1.6$ \citep[see, e.g.,][]{beelen06}, these flux densities correspond to IR luminosities (integrated between 8--1000\,mm) of $(1.3-2.6)\times 10^{12}$\,\Lsun{}, $(2.8-5.4)\times 10^{11}$\,\Lsun{}, and $(1.5-3.0)\times10^{11}$\,\Lsun{} for the three components, where the range refers to dust temperatures spanning between 35 and 45\,K. Under the assumption that the dust is optically thin, and following the normalization by \citet{dunne00}, we derive dust masses of $(1.4-2.6)\times 10^8$\, $(2.1-4.4)\times10^7$, and $(1.5-3.1)\times10^7$\,\Msun{} for Q, E, and W, respectively. We note that a typical gas--to--dust ratio of 100 \citep[e.g.,][]{berta16} would yield a significantly lower gas mass than those based on \Cii{} derived via the \citet{zanella18} calibration, possibly due to the caveats in the adoption of a single value for $\alpha_{\rm [CII]}$, and due to the limited surface brightness sensitivity of our observations. As no spatially--resolved constrain on $T_{\rm dust}$ is available, in the following we assume $T_{\rm dust}$=35\,K everywhere in the system.

\begin{figure}
\includegraphics[width=0.49\textwidth]{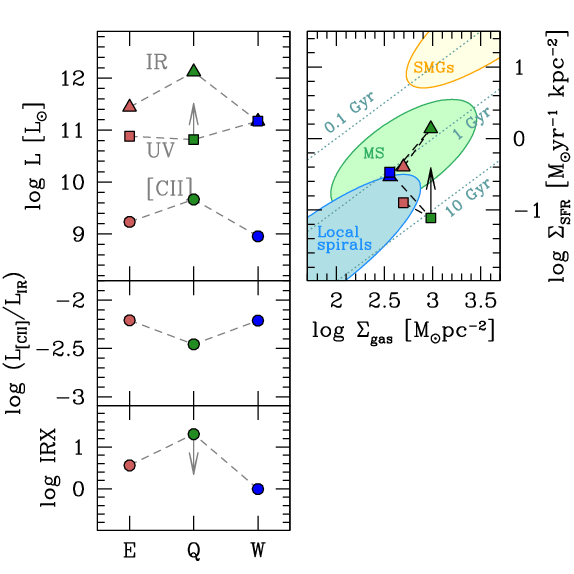}
\caption{ISM and SFR diagnostics in PJ308--21. {\em Left}: IR, UV, \Cii{} luminosities and their ratios in Q, E, W, derived from the apertures shown in Fig.~\ref{fig_hst_maps}. The values derived for both the quasar host galaxy and the companion blobs are comparable to those in typical star-forming high--redshift galaxies. {\em Right}: The star formation vs.\ gas surface density, or ``star--formation law'' in Q, E, and W. SFRs from UV (squares) and IR (triangles) are shown as a function of gas surface densities derived based on the \citet{zanella18} calibration from \Cii. The loci at constant depletion times are shown, as well as the range of values observed in SMGs, in local spirals, and in global measurements of main sequence galaxies \citep[see, e.g.,][]{hodge15}. All the components of PJ308--21 appear in line with the typical values of main sequence galaxies at high redshift.}
\label{fig_diagnostics}
\end{figure}

The \Cii{}/IR luminosity ratio is a commonly used ISM diagnostic, with values around $3\times10^{-3}$ for local star forming galaxies, and $<3\times10^{-4}$ in compact starbursts and ULIRGs \citep[e.g.,][]{herreracamus15,diazsantos17}. The \Cii{}/IR ratio in PJ308--21 shows a wide range of \Cii{}/IR values, down to $3\times10^{-4}$ close to the quasar location. Once averaged over the apertures shown in Figs.~\ref{fig_alma_vel}--\ref{fig_hst_maps}, the \Cii{}/IR in Q is $3.5\times 10^{-3}$, i.e., 2$\times$ lower than in E and W (see Fig.~\ref{fig_diagnostics}). For a higher $T_{\rm dust}$=45\,K in the regions close to the quasar, where the gas and star formation surface densities are highest (see below), the \Cii{}/IR luminosity ratio drops by a factor $\sim 2$.

\subsection{Star formation rate surface density}

Our {\em HST} image of PJ308--21 probes the rest-frame UV starlight from young stars, and therefore traces the unobscured component of star formation\footnote{The diffuse rest-frame UV emission observed with {\em HST} can also be attributed (at least in part) to dust-scattered light from the quasar itself \citep[see, e.g.,][]{zakamska06}. At present, we cannot unambiguously distinguish between the two scenarios; however, we point out that the UV--brightest knots in E, W, and in the North-West component are associated with relatively lower dust surface brightness, contrary to a simple reflection scenario. We will ignore the impact of reflected light in the remainder of our analysis.}. Complementarily, the ALMA dust continuum reveals star formation that is enshrouded by dust. The combination of the two is thus a proxy of the total star formation in this system. After PSF subtraction of the {\em HST} image, and after masking the central 2.5\,kpc ($\approx 0.5''$, dominated by residuals; this area is also the most sensitive to color terms in the PSF model), we measure F140W magnitudes of 25.33, 25.17, and 24.44 mag for Q, E, and W, yielding a rest--frame UV luminosity of $\log \nu L_\nu$(1900\,\AA) [\Lsun] = 10.82, 10.88, and 11.17, respectively. By construction, this is only a lower limit on the UV emission of Q, due to the uncertainties in removing the nuclear emission and the masking of the central pixels. Following \citet{kennicutt12}, these luminosities translate into UV--based SFRs of 11, 13, and 25\,\Msun\,yr$^{-1}$, respectively. The obscured SFRs are derived from IR luminosities following \citet{kennicutt12}: 290, 60, and 32 \Msun{}\,yr$^{-1}$ for Q, E, and W, assuming $T_{\rm dust}$=35\,K. For $T_{\rm dust}$=45\,K, our estimates of the IR--based SFRs would roughly double.

The IR--to--UV luminosity ratio, or ``IR excess'', IRX, is a proxy of the relevance of obscured--to--unobscured SFR, and can be used to study dust reddening at high $z$ \citep[e.g.][]{whitaker14,bouwens16,wang18}. We find that both E and W have IRX values consistent with those of typical low--extinction high--$z$ galaxies (\citealt{whitaker14}; see Fig.~\ref{fig_diagnostics}). 

We estimate the SFR surface density, $\Sigma_{\rm SFR}$, from the full--resolution IR-- and UV--based SFRs. We find the highest value of $\Sigma_{\rm SFR}$=14\,\Msun{}\,yr$^{-1}$\,kpc$^{-2}$ (set by the 4.0\,kpc$^2$ area of the ALMA beam in our observations) for the IR--based SFR at the position of the quasar. This is well below the Eddington limit \citep[$\sim1000$ \Msun{}\,yr$^{-1}$\,kpc$^{-2}$; see, e.g.,][]{walter09,hodge15} even if we assume a modest IRX$\approx$1 to account for the (unconstrained) contribution of the unobscured SFR at this position, or for a higher $T_{\rm dust}$. However, we stress that these estimates are based on average emission over relatively large apertures (a few kpc$^2$ in area), and much higher values could be in place on local scales.

In Fig.~\ref{fig_diagnostics}, {\em right}, we compare the average $\Sigma_{\rm SFR}$ values estimated over the apertures shown in Figs.~\ref{fig_alma_vel}--\ref{fig_hst_maps} with the gas surface density derived from \Cii{}, $\Sigma_{\rm gas}$, via the \citet{zanella18} calibration. 
We find that $\Sigma_{\rm SFR}$ and $\Sigma_{\rm gas}$ in PJ308--21 are in line with the values typically observed in global observations of main sequence galaxies at $z$=1--3 \citep{tacconi13}, and are significantly lower than the values observed in intense starbursts and SMGs at high redshifts (e.g., \citealt{hodge15}; \citealt{chen17}). The average depletion time in the system is $\sim 1$\,Gyr, i.e., if no significant gas accretion occurs, this system is expected to run out of fuel for star formation by $z\sim4$.

\subsection{Dynamics of the system}

The morphology and the complex velocity structure observed in PJ308--21 (see Fig.~\ref{fig_alma_vel}) appear inconsistent with an interpretation in terms of gas expanding from the center outward in response to the feedback from the quasar. In particular, the opposite signs of the projected velocity and velocity gradients in E and W, as well as the low gas velocity dispersion in E ($\lsim 100$\,\kms{} along the line of sight, compared to $>700$\,\kms{} values associated with outflows in, e.g., J1152+5251 by \citealt{cicone15}), defy basic expectations for an outflow scenario. 

Conversely, in this section we test whether a toy model of the tidal disruption of a single satellite galaxy in close interaction with the quasar host can account for the observed gas dynamics in PJ308--21. This simplistic approach is not a fit to the data, but rather a proof of concept that the tidal disruption scenario works for this system. Specifically, we test whether the proposed dynamical description of PJ308--21 succeeds in predicting the  observed range of line--of--sight gas velocity, and the spatial extent of PJ308--21. 
\begin{figure}
  \includegraphics[width=0.49\textwidth]{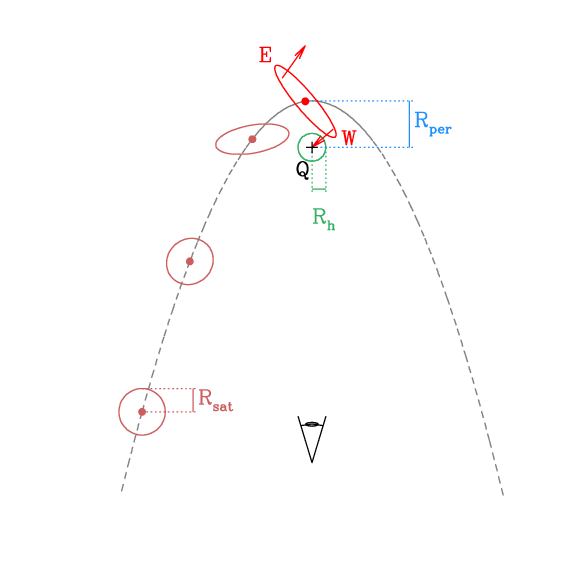}\\
  \caption{Sketch of the toy model. The satellite galaxy (dots with ellipses) approaches the quasar host galaxy (Q, marked by a +, and with a scale radius $R_{\rm h}$) via a highly--eccentric parabolic orbit (grey, dashed line). The system is observed when the satellite is close to the pericenter ($R_{\rm per}$). The satellite is tidally stretched along its course, thus leading to the wide range of projected velocities and the spatial extent between the eastern and western wings of the satellite (E and W). We stress that this cartoon is not in scale.}
\label{fig_cartoon}
\end{figure}

Strong tidal perturbations should arise when the mass ratio between the satellite and the host is $M_{\rm sat}/M_{\rm host}<(R_{\rm sat}/R_{\rm peri})^3$, where $R_{\rm sat}$ is the satellite's scale size, and $R_{\rm peri}$ is the orbital pericenter. In this regime, both the velocity gradient and the spatial stretch observed between E and W would be due to the tidal interaction with Q. We sketch a cartoon of the model in Fig.~\ref{fig_cartoon}.

We assume a parabolic, highly eccentric orbit. We also postulate that our observations have caught the satellite close to the pericenter, as suggested by the high magnitudes and different signs of the line--of--sight velocities of E and W. Thus, the current satellite velocity ($v_{\rm sat,peri}$) equals the escape velocity: $v_{\rm sat,peri}\approx v_{\rm esc}=\sqrt{2\phi(R_{\rm peri})}$. The potential of the host\footnote{We here implicitly assume spherical symmetry, although the estimate of the escape velocity is insensitive to changes in axis ratio.}, $\phi$, at a distance $R$, is the sum of the baryonic potential $\phi_{\rm bar}=G M_{\rm host,bar}/R$ and of the Navarro Frenk and White profile dark matter (DM) potential, $\phi_{\rm DM}$, which can be expressed in terms of the enclosed mass $M_{\rm  host,DM}(< R)$ and scale radius $R_{h}$ as:
\begin{equation}\label{eq_phi}
  \phi_{\rm DM}(R)=\frac{G}{R}M_{\rm host,DM}(< R)\left[1-\frac{R/(R+R_h)}{\ln(1+R/R_h)}\right]^{-1}.
\end{equation}
The enclosed DM mass is estimated from the observed velocity curve of the primary galaxy, Q:
\begin{equation}\label{eq_MhostDM_inR}
  M_{\rm  host,DM}(< R)=\frac{(\alpha v_{\rm host, circ})^2 R}{G}-M_{\rm host,bar},
\end{equation}
where $v_{\rm host, circ}$ is the observed circular velocity, $\alpha \ge 1$ is used to account for possible observational underestimates due to, e.g., beam smearing and line--of--sight projection \citep[see, e.g.,][]{lupi19}. The DM mass fraction within $R$ (compared to the total) is:
\begin{equation}\label{eq_MhostDM_ratio}
\frac{M_{\rm host,DM}(<R)}{M_{\rm host,DM}}= \frac{\ln(1+R/R_h)-R/(R+R_h)}{\ln(1+c)-c/(1+c)},
\end{equation}
where $c$=$R_{\rm vir}/R_h$ is the concentration parameter of the halo, $R_{\rm vir}$ is its virial radius, and we assume $c$=10.

\begin{figure}
  \includegraphics[width=0.49\textwidth]{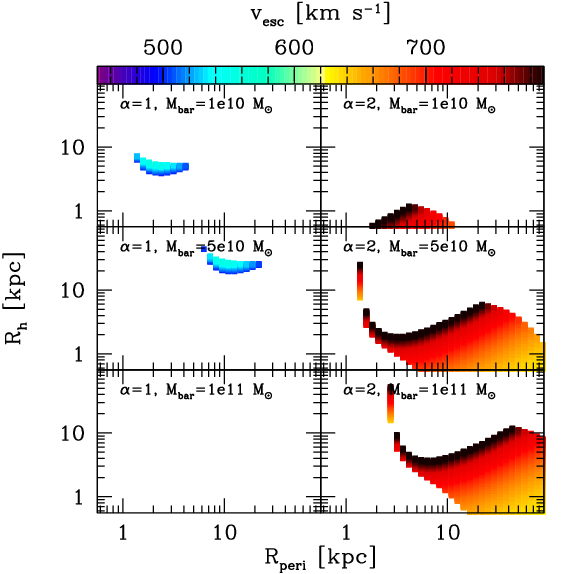}\\
  \caption{Maps of escape velocity $v_{\rm esc}(R_{\rm peri})$ as function of $R_{\rm peri}$ and $R_h$, for $\alpha v_{\rm host,circ}(R_{\rm peri})$=200\,\kms{} (left) or 400\,\kms{} (right), and for a baryonic mass of the quasar host within the pericenter, $M_{\rm host,bar}(<R_{\rm peri})$, equal to $(0.1, 0.5, 1)\times 10^{11}$ \Msun{} (top to bottom). The observed kinematics in PJ308--21 appear consistent with the tidal interaction of a single satellite galaxy by the more massive quasar host galaxy, within a broad range of input parameters. No solution consistent with the expected range of $v_{\rm esc}$ is found for $\alpha=1$ and $M_{\rm host,bar}(<R_{\rm peri})= 10^{11}$\,\Msun, since such a mass would imply a larger circular velocity.}
\label{fig:3d}
\end{figure}

Under these assumptions, the escape velocity is fully determined by a combination of ($M_{\rm host}$, $R_{\rm peri}$, $R_h$, $\alpha$). Fig.~\ref{fig:3d} shows the parameter combinations that yield escape velocities in broad agreement with the observed velocity differences between Q, E, and W. We only consider cases that yield a total DM--to--baryon mass ratio of the quasar host $3<M_{\rm host,DM}/M_{\rm host,bar}<30$.

We can now use the sub-set of input parameter values that yield an escape velocity consistent with the observations to infer the expected spatial stretch of the satellite. Using equation \ref{eq_MhostDM_ratio}, we can infer the mass ratio between the innermost and outer parts of the satellite close to the pericenter. This is then used to infer the relative velocity of the two sides of the galaxy, $\Delta v$, via equations \ref{eq_phi} and \ref{eq_MhostDM_inR}. The timescale of the interaction is set by $\Delta t=R_{\rm peri}/v_{\rm peri}$. The resulting spatial stretch is thus $\Delta R\approx \Delta v \Delta t$. This scaling successfully explains the size of PJ308--21. E.g., if we assume $R_{\rm per}$=10\,kpc, an initial satellite radius $R_{\rm sat}$=5\,kpc, a scale radius of the quasar host of $R_{\rm h}$=3\,kpc, $\alpha$=2, and a host mass of $10^{11}$\,\Msun{}, we obtain $\Delta v\approx 240$\,\kms{}, $\Delta t\approx 45$\,Myr, and a major semi-axis of the satellite of $\Delta R\approx 11$\,kpc.

A single satellite scenario is thus consistent with the observations, either requiring a very diffuse secondary scattering onto a more massive primary, or a more compact secondary undergoing a close flyby (down to $R_{\rm peri}\approx 1$ kpc). Such a scenario naturally accounts for the low velocity dispersion of \Cii{} in E (the main tidal feature) and the higher value in W (as the tidal disruption creates a bridge between the satellite and the quasar host that is roughly aligned with the line of sight). A primary baryonic mass of $M_{\rm host,bar}>10^{11}$\,\Msun{} does not result in any solution consistent with the data if the observed velocity along the line of sight is used as a good proxy for the circular velocity. Although this can hint to a possible tension between the dynamical constraints and the \Cii{}--based estimate of the gas mass, we stress that small variations of the model, e.g., including non-rotational components in the quasar host \citep[e.g.][]{lupi19}, the rotation of the secondary before the pericenter, or gas-dynamical effects in close peri-passages \citep{barnes02,capelo17,blumenthal18} could modify the gas velocity map by up to $\sim 100$ km s$^{-1}$, allowing for a slightly larger baryonic mass of the primary.

\section{Discussion \& Conclusions}\label{sec_discuss}

We present new kpc--scale ALMA and {\em HST} imaging of the quasar PJ308--21 at $z$=$6.2342$. We find extended emission from young stars, dust, and gas (traced via the \Cii{} 158$\mu$m line) stretching both eastward and westward of the quasar host for a total projected extent exceeding 20\,kpc. The system has a complex velocity structure covering $>500$\,\kms{} both blue-- and redward of the quasar systemic redshift. The close morphological match between rest--frame UV light, dust, and \Cii{}--traced gas unambiguously associate the extended emission with the immediate environment of the quasar, and rules out a foreground projected object. The morphology, size, velocity, and velocity dispersion structure of the system are reminiscent of local gas--rich mergers \citep[e.g.,][]{tacconi99}. The luminosities of all the components, and their ratios, are consistent with values typically observed in high--redshift galaxies, and do not seem to reproduce the shock--heated values observed in outflows or in shock fronts \citep[e.g., $L_{\rm [CII]}/L_{\rm IR}>0.01$, see][]{appleton13}. We demonstrate that the observed velocity range and spatial extent can be accounted for by a simple model of a tidally--disrupted satellite galaxy in close encounter with the quasar host galaxy. 

In summary, PJ308--21 is one of the earliest mergers imaged (in terms of cosmic time), and represents a unique laboratory to study the assembly of massive galaxies at cosmic dawn. The quasar host galaxy and its surroundings are natural test cases for studies of other ISM tracers (e.g., molecular lines such as CO or H$_2$O, far-infrared fine structure lines, dust continuum) using ALMA. Moreover, this system is the first, unambiguous example of a merging quasar host galaxy at $z>6$, and the first case of a quasar host galaxy stellar light detected in its rest-frame UV emission at these redshifts. This makes PJ308--21 a prime target for investigations with the James Webb Space Telescope, which will enable a direct measurement of the already--assembled stellar mass of this unique system.

\acknowledgements

We thank the referee for their excellent feedback on the manuscript. We thank the ALMA Director for granting time for the DDT observations presented here. F.W., Ml.N., Ma.N., and B.P.V.{} acknowledge support from the ERC Advancedgrant 740246 (Cosmic\_Gas). D.R. acknowledges support from the National Science Foundation under grant number AST-1614213.
%FW, BPV, EPF acknowledge support through ERC grant Cosmic\_Dawn. DR acknowledges support from the National Science Foundation under grant number AST-1614213 to Cornell University. 
\facility{ALMA} data: 2015.1.01115.S and 2016.A.00018.S. ALMA is a partnership of ESO (representing its member states), NSF (USA) and NINS (Japan), together with NRC (Canada), NSC and ASIAA (Taiwan), and KASI (Republic of Korea), in cooperation with the Republic of Chile. The Joint ALMA Observatory is operated by ESO, AUI/NRAO and NAOJ. 
\facility{HST}. Based on observations made with the NASA/ESA Hubble Space Telescope, obtained from the Data Archive at the Space Telescope Science Institute, which is operated by the Association of Universities for Research in Astronomy, Inc., under NASA contract NAS 5-26555. These observations are associated with program 14876. Support for this work was provided by NASA through grant number 10747 from the Space Telescope Science Institute, which is operated by AURA, Inc., under NASA contract NAS 5-26555.

\label{lastpage}

\end{document}